
\def\etal{{\it et~al.}}

\def\halfspace{\baselineskip=12pt plus .1pt}

\def\papersize{\magnification=1200}  

\font\tfont=cmmib10
\newfam\vecfam

\textfont\vecfam=\tfont \scriptfont\vecfam=\seveni
\scriptscriptfont\vecfam=\fivei

\parindent=40pt
\settabs 7 \columns
\tolerance=1600
\parskip 1ex
\def\foolit{\ifnum\pageno > 1 \number\pageno\fi}
\raggedbottom
\def\frac#1#2{{#1\over#2}}
\def\Mesz{M\'esz\'aros\ }
\def\Pacz{Paczy\'nski\ }

\def\ctl{\centerline}

\def\gbr{\goodbreak\noindent}

\def\ref{\par \noindent \hangindent=2pc \hangafter=1 }
\def\etal{{\it et~al.\ }}
\def\mathnew{\mathsurround=0pt}
\def\simov#1#2{\lower .5pt\vbox{\baselineskip0pt \lineskip-.5pt
	\ialign{$\mathnew#1\hfil##\hfil$\crcr#2\crcr\sim\crcr}}}
\def\simg{\mathrel{\mathpalette\simov >}}
\def\siml{\mathrel{\mathpalette\simov <}}
\def\lambdabar{\mathrel{\lower 1pt\hbox{$\mathchar'26$}\mkern-9mu
        \hbox{$\lambda$}}}

\def\cm{{\rm\,cm}}

\def\bsk{\vskip 3ex\noindent}

%

\def\cm{~\rm{cm}}

\def\s{~\rm{s}}

\def\cmcui{~ {\rm cm}^{-3} }

\def\MeV{~\rm{MeV}}
\def\GeV{~\rm{GeV}}

\def\ergs{~\rm{ergs}}

\def\G{~\rm{G}}

\def\Hz{~\rm{Hz}}

\papersize
\halfspace
%
$~$\bsk
\ctl{\bf DELAYED GeV EMISSION FROM COSMOLOGICAL GAMMA-RAY BURSTS :}
\ctl{\bf Impact of a Relativistic Wind on External Matter}
\bsk
\ctl{ P. \Mesz$^{1}$ and M.J. Rees$^{2}$}
\bsk
\ctl{$^1$~ Pennsylvania State University, 525 Davey Lab, University Park, PA
16803}
\ctl{$^2$~ Institute of Astronomy, Madingley Road, Cambridge CB3 OHA, England}
\bsk
\ctl{ Submitted to M.N.R.A.S.: ~4/21/94;~ accepted:~~~~~~~~~~~~ }
\bsk
\ctl{\bf Abstract}
\bsk
Sudden collapse of a compact object, or coalescence of a compact binary, can
generate an unsteady relativistic wind that lasts for a few seconds. The wind
is
likely to carry a high magnetic field; and its Lorentz factor depends on the
extent to which it is 'loaded' with baryons. If the Lorentz factor is $\sim
100$,
internal dissipation and shocks in this wind produce a  non-thermal gamma-ray
burst, detectable in the range $0.1\MeV \siml E_\gamma \siml 0.1-1\GeV$ out to
cosmological distances.
The cooled wind ejecta would subsequently be decelerated by the external
medium.
The resultant blast wave and reverse shock can then give rise to a second burst
component, mainly detectable in the GeV range, with a time delay relative to
the MeV burst ranging from minutes to hours.
\bsk\bsk
\ctl{ \bf 1.~ Introduction}
\bsk
Emission at energies in the GeV range has been recently reported in a
gamma-ray burst (GRB) from EGRET observations (Mukherjee, \etal, 1994),
continuing to arrive up to $\sim 1$ hour after the MeV-band emission detected
by BATSE whithin 0.5$^o$ of the GeV location. Previous reports of GeV emission,
e.g. from the `Superbowl' burst (Dingus, \etal, 1994) also showed a (relatively
shorter) delay respect to the MeV burst. A `simple' fireball model would have
problems in explaining these delayed bursts, whether galactic or cosmological,
because most of the energy would escape in a single, very short burst when the
fireball becomes optically thin. In previous papers we
have explored various `non-simple' fireball and wind models which avoid the
problems encountered in the pre-1992 simple models (\Mesz and Rees, 1992, 1993;
Rees and \Mesz, 1994; see also Narayan, \etal, 1992, Woosley,
1993, Usov, 1994, Katz, 1994, \Pacz and Xu, 1994, Thompson, 1994).
Here we show that a dissipative unsteady wind model which is decelerated by an
external medium provides a natural explanation for having a classical MeV band
GRB of the usual duration ($10^{-1}-10^3\s$), followed by a harder GeV burst
which is spread out or delayed over intervals ranging from minutes to hours,
with duration comparable to the delay. The MeV burst is produced by internal
shocks in the unsteady relativistic wind produced by a primary event,
lasting seconds (e.g. Rees and \Mesz, 1994; see also \Pacz and Xu, 1994,
where neutrino and radio emission are discussed). The GeV burst occurs
when the cooled, coasting baryons from the wind run into the external
interstellar medium (ISM), producing a blast wave moving ahead and a reverse
shock moving back into the ejecta (e.g. \Mesz and Rees, 1993). For the
baryon loading factors producing a classical MeV burst, and an external
ISM of standard density, the relative delay or spread of the GeV emission
respect to the BATSE MeV-band emission extends up to $\sim$ hours.
\bsk\gbr
\ctl{\bf 2.~ Internal Shocks}
\bsk
In an unsteady relativistic wind of duration $t_w$ with an energy input varying
at
the base on a timescale $t_{v}$, a relatively large baryon loading factor
corresponding to $\eta=(L/{\dot M}c^2)\simg \eta_{min}\sim 3\times 10^1
(L_{51}/t_{v})^{1/5}$ leads to internal shocks outside the wind photosphere
which
dissipate a sizable fraction of the bulk kinetic energy of the wind (Rees and
\Mesz,
1994). If the Poynting flux provides a fraction $\alpha$ of the total
luminosity $L$ of the wind, the comoving magnetic field at the dissipation
radius $r_{dis}$ where the internal shocks occur is
$B_{dis}\sim 10^4\alpha^{1/2}L_{51}^{1/2}t_{v}^{-1}\eta_2^{-3}\G$,
leading to a high radiative efficiency even for $t_{v}$ as long as seconds.
The internal shocks are marginally relativistic, since blobs in
the wind are moving at alomost the same speed.
For electrons accelerated in the internal shocks to a minimum Lorentz factor
$\gamma_m\sim \kappa \sim 10^2\kappa_2$ (where $1 \siml \kappa \siml
m_p/m_e$)the observed synchrotron break would come at a frequency
$$
\nu_{sy,ob}\sim 10^{16}\alpha^{1/2}L_{51}^{1/2}\kappa_2^2 t_{v}^{-1}
                                          \eta_2^{-2}~\Hz~.
                                                                    \eqno(1)
$$
Inverse Compton (IC) losses would also be important, the ratio of the
synchrotron photon energy to the magnetic field energy being $u_{sy}/u_B\sim
\rho(\sigma_T/m_p) \gamma_m^2 f r_{dis} \eta^{-1} \sim 5\times 10^1 L_{51}
f \kappa_2^2 t_v^{_1}\eta_2^{-5}$, where $f\siml 1$. The IC scattered
synchrotron
photons would have a break frequency of
$$
\nu_{IC,ob}\sim \nu_{sy,ob}\gamma_m^2 \sim 10^{20} \alpha^{1/2}L_{51}^{1/2}
\kappa_2^4 t_{v}^{-1} \eta_2^{-2} ~\Hz~.\eqno(2)
$$
This would, therefore, give a burst of duration $t_w$ in the MeV range,
detectable by BATSE at cosmological distances. The compactness parameter in the
wind rest frame, depending on the energy and the radius at which the
dissipation
shocks occur, becomes significant for $\simg 0.1-1~\GeV$, so photo-pair
creation
would strongly supress this burst over the higher part of the EGRET band.
\bsk\gbr
\ctl{\bf 3.~ External Shocks}
\bsk
The wind ejecta, after having produced the MeV burst through internal
dissipation
shocks, will continue coasting. By the time it is decelerated by encounter with
the external medium at a radius $\sim 10^{17}\cm$, the wind ejecta can be
treated
as an `impulsive' cooled fireball. (The thickness of the `shell' that
would have developed from an impulsive calculation with $\eta\sim 10^2$
by the time it gets out to $\sim 10^{17}\cm$, e.g. \Mesz, Laguna and Rees,
1993,
is larger than the few light-seconds in the lab frame that results from the
extended duration of the wind). This is compatible with the observation that
the duration of the primary (MeV) burst is generally $t_b\siml 10^2-10^3\s$,
while the delay and duration of the GeV burst is generally longer than this.
As the ejecta is decelerated, a relativistic blast wave moves into the ISM
ahead,
while a reverse shock propagates into the ejecta. If the wind was dominated by
the Poynting flux, the magnetic field remains in equipartition as long as the
ejecta is in the wind regime, and the ratio of magnetic to rest mass energy
density drops as $r^{-1}$ thereafter. Comparing the MeV and GeV burst
durations,
the field would be at most a factor $\sim 10^{-1}$ below equipartition in
the deceleration shocks. Alternatively, if turbulent field growth behind the
shocks leads to the latter having a fraction $\lambda$
of the equipartition energy, the comoving field is $B_{dec}\sim 10^1
n_0^{1/2}\eta_2\lambda^{1/2}\G$. The blast wave would produce electrons with
a minimum Lorentz factor $\gamma_m\sim \kappa\eta \sim 10^4\kappa_2\eta_2$.
For an external medium (ISM) of density $n_{ext}\sim 1~n_0\cmcui$
the observer-frame synchrotron frequency break would be at a frequency
$$
\nu_{sy,ob}\sim 10^{17}n_{0}^{1/2}\lambda^{1/2}\kappa_2^2 \eta_2^4 ~\Hz~.
                                                        \eqno ( 3)
$$
The observed IC scattered synchrotron photons in the blast wave zone would
have a break frequency of
$$
\nu_{IC,ob}\sim \nu_{sy,ob}\gamma_m^2 \sim 10^{25}n_{0}^{1/2}\lambda^{1/2}
\kappa_2^4 \eta_2^6~\Hz ~. \eqno(4)
$$
The energy loss is dominated by the IC process, so most of the deceleration
energy liberated by the external shock comes out at frequencies near (4). This
is similar to the calculations for the piston model (Fig. 7) or the turbulent
model (Fig. 5) of \Mesz, Rees and Papathanassiou, 1994.
(However, we note that the observations may not require the MeV fluence to be
low,
since the BATSE trigger is activated by exceeding a certain photon count {\it
rate};
a long ($t\simg 10^3\s$) burst could in principle have significant MeV fluence
without triggering BATSE).
The deceleration occurs at a radius $r_{dec}\sim 5\times 10^{16} n_{0}^{-1/3}
E_{51}^{1/3}\eta_2^{-2/3}\cm$, and it occurs after a time delay
$$
t_{del} \sim r_{dec}/(c \eta^2)\sim 5\times 10^2 n_{0}^{-1/3}E_{51}^{1/3}
                             \eta_2^{-8/3}~\s~.\eqno(5)
$$
The total duration of this burst is comparable to $t_{del}$ if the external
medium
is smooth, and for a narrow range of $\eta=(L/{\dot M}c^2)$ corresponding to
$t_{del} > t_w$ the external shock would produce a GeV burst delayed by
$t_{del}$
respect to the dissipation shock burst. However for a wider range of $\eta$
or an inhomogeneous external medium,
the beginning of the external shock burst could overlap with the later stages
of the
internal shock burst. If the external medium were blobby on scales $r_{blob}<
r_{dec}$, time structure of order $t_{str}\sim r_{blob}c^{-1} \eta^{-2}$ would
be
observed. Photon statistics would also affect the time structure at GeV
energies.
\bsk\gbr
\ctl{\bf 4.~ Discussion}
\bsk
Second generation (`non-simple') cosmological wind and fireball models,
such as previously proposed by us, have been shown to be able to explain all
the major features of classical GRBs, in particular the 0.1-100 MeV
non-thermal spectrum, the total energy, duration and time structure.
In this paper, we have discussed how these models also provide a natural
explanation for the longer lasting GeV emission observed by EGRET in some
classical GRBs.

In the model proposed here, the MeV burst is observed first, and is caused
by a `primary event' producing a (possibly magnetized) wind of luminosity
$L\sim 10^{51}L_{51}\ergs$ and duration $t_w\sim 10^{-1}-10^3\s$. The average
baryon loading may be relatively high, leading to bulk Lorentz factors
$\Gamma\sim \eta=(L/{\dot M}c^2) \sim 10^2$. Intrinsic time variabity
of the energy input or loading factor leads to blobs or shells of slightly
different
Lorentz factors colliding and dissipating an appreciable fraction of the bulk
kinetic energy of the wind (e.g. Rees and \Mesz, 1994; also see \Pacz, 1991,
Thompson, 1994, \Pacz and Xu, 1994). A large baryon loading, and a weakly
emitting quasi-thermal wind photosphere are acceptable, because the MeV burst
occurs in dissipative shocks {\it outside} the photosphere, at
the expense of the baryon bulk kinetic energy (Rees and \Mesz, 1994), in an
optically thin environment leading to a nonthermal MeV spectrum. Due to
the photon compactness, this burst would be supressed above about $0.1-1\GeV$.

A second burst arises when the wind ejecta, carrying a fraction of the
initial baryon kinetic energy, has swept up an amount of external mass
$M_{ext}\sim \Gamma^{-1}M_{ej}$, e.g. \Mesz and Rees, 1993.
The external medium could be either the interstellar medium, or material
ejected
by the progenitor star (possibly in the form of a shell) in a mass loss phase
prior to the GRB event. For a standard ISM of density 1/c.c. and $\eta\sim
10^3$, the duration of this burst would be of order $t_{dec}\sim t_{del}\sim$
seconds, and the spectrum extends from MeV to GeV energies (e.g. \Mesz, Rees
and Papathanassiou, 1994). For higher baryon loading, e.g.  $\eta \siml 10^2$,
the duration of this burst can exceed $\simg 10^3\s$ and the spectrum peaks at
higher energies, in the GeV range (although for such durations even
a significant MeV fluence might go undetected).

Depending on the baryon loading of the ejecta and the density or blobbiness of
the external medium, the time-delay of the GeV burst relative to the
previous MeV burst will be in the range $5\s \siml t_{del} \siml 5\times
10^3\s$,
and the burst lasts some fraction of this value. For a range of $\eta$, the two
bursts overlap, the GeV emission lasting longer. These bursts would be visible
from cosmological distances, and would have fluxes, spectra, duration and rate
of
occurrence compatible with the observations.
Weak X-ray and optical signatures from these bursts (\Mesz, Rees and
Papathanassiou,
1994) may also be detectable by next generation omnidirectional space
detectors.
\bsk
{\it Acknowledgements}: We are grateful to C. Fichtel, G. Fishman and R.
Mukherjee
for timely comments. This research has been partially supported through NASA
NAGW-1522, NAG5-2362 and by the Royal Society.
\bsk\gbr
\ctl{\bf References}
\bsk
\ref Dingus, B., \etal, 1994, in {\it Proc. Huntsville Gamma-ray Burst Wkshp.},
  eds. G. Fishman, K. Hurley, J. Brainerd, (AIP: New York), in press
\ref Katz, J.I., 1994, Ap.J., 422, 248
\ref \Mesz, P. and Rees, M.J., 1992, Ap.J., 397, 570
\ref \Mesz, P. and Rees, M.J., 1993, Ap.J., 405, 278
\ref \Mesz, P., Laguna, P. and Rees, M.J., 1993, Ap.J., 415, 181.
\ref \Mesz, P., Rees, M.J. and Papathanassiou, H., 1994, Ap.J., in press
\ref Mukherjee, R., \etal, 1994, paper at the Washington, D.C., APS Spring
Meeting.
\ref Narayan, R., Paczynski, B. and Piran, T., 1992, Ap.J.(Letters), 395, L83
\ref Paczy\'nski, B., 1991, Acta Astronomica, 41, 257
\ref Paczy\'nski, B. and Xu, G., 1994, Ap.J., in press
\ref Rees, M.J. and \Mesz, 1994, Ap.J.(Letters), submitted
\ref Thompson, C., 1994, M.N.R.A.S., in press
\ref Woosley, S., 1993, Ap.J., 405, 273
\end